\documentclass[final,5p,times,twocolumn]{elsarticle}

\usepackage{graphicx}
\usepackage{subfigure}
\usepackage{multirow}
\usepackage{wrapfig}
\usepackage{array}

\usepackage{tabularray}
\usepackage{tabularx}
\usepackage{rotating}
\usepackage{makecell}
\usepackage{booktabs}
\usepackage{csquotes}

\usepackage[utf8]{inputenc}
\usepackage{amsmath}
\usepackage{amsthm}
\usepackage{amsfonts}
\usepackage{epsfig}
\usepackage{psfrag}
\usepackage{pstricks}
\usepackage{algorithm}
\usepackage{stfloats}

\usepackage{cuted}

\usepackage{algorithmicx}
\usepackage{algpseudocode}  
\floatname{algorithm}{Algorithm}

\graphicspath{{./Figures/}}

\usepackage{amssymb}
\usepackage{bm}
\usepackage{hyperref}
\hypersetup{
            colorlinks=true,
            linkcolor=blue,
            anchorcolor=blue,
            citecolor=blue
            }

\biboptions{comma,sort&compress}

\usepackage{enumitem}
\setlist{nosep,topsep=-\parskip}

\usepackage{ulem}

\journal{VI}

\begin{document}

\begin{frontmatter}

\title{Intelligent CAD 2.0}
\author[]{Qiang Zou\corref{cor}}\ead{qiangzou@cad.zju.edu.cn}
\author[]{Yincai Wu}
\author[]{Zhenyu Liu}
\author[]{Weiwei Xu}
\author[]{Shuming Gao}

\cortext[cor]{Corresponding author.}
\address{State Key Laboratory of CAD$\&$CG, Zhejiang University, Hangzhou, 310058, China}

\begin{abstract}
Integrating modern artificial intelligence (AI) techniques, particularly generative AI, holds the promise of revolutionizing computer-aided design (CAD) tools and the engineering design process. However, the direction of ``AI+CAD" remains unclear: how will the current generation of intelligent CAD (ICAD) differ from its predecessor in the 1980s and 1990s, what strategic pathways should researchers and engineers pursue for its implementation, and what potential technical challenges might arise?

As an attempt to address these questions, this paper investigates the transformative role of modern AI techniques in advancing CAD towards ICAD. It first analyzes the design process and reconsiders the roles AI techniques can assume in this process, highlighting how they can restructure the path humans, computers, and designs interact with each other. The primary conclusion is that ICAD systems should assume an intensional rather than extensional role in the design process. This offers insights into the evaluation of the previous generation of ICAD (ICAD 1.0) and outlines a prospective framework and trajectory for the next generation of ICAD (ICAD 2.0).
\end{abstract}




\end{frontmatter}


\section{Introduction}
\label{sec:intro}
Starting from Sutherland's Sketchpad~\cite{sutherland1964sketch} in the 1960s, computer-aided design (CAD) techniques have evolved significantly, from 2D drafting tools in the 1970s to 3D surface and solid modeling systems in the 1980s, and later to feature-based parametric design software in the 1990s and 2000s~\cite{shah1995parametric}. CAD has since become extensively adopted across various industries, including mechanical, aeronautical, electrical, and architectural engineering.

\begin{figure*}[t]
\centering
\includegraphics[width=0.8\textwidth]{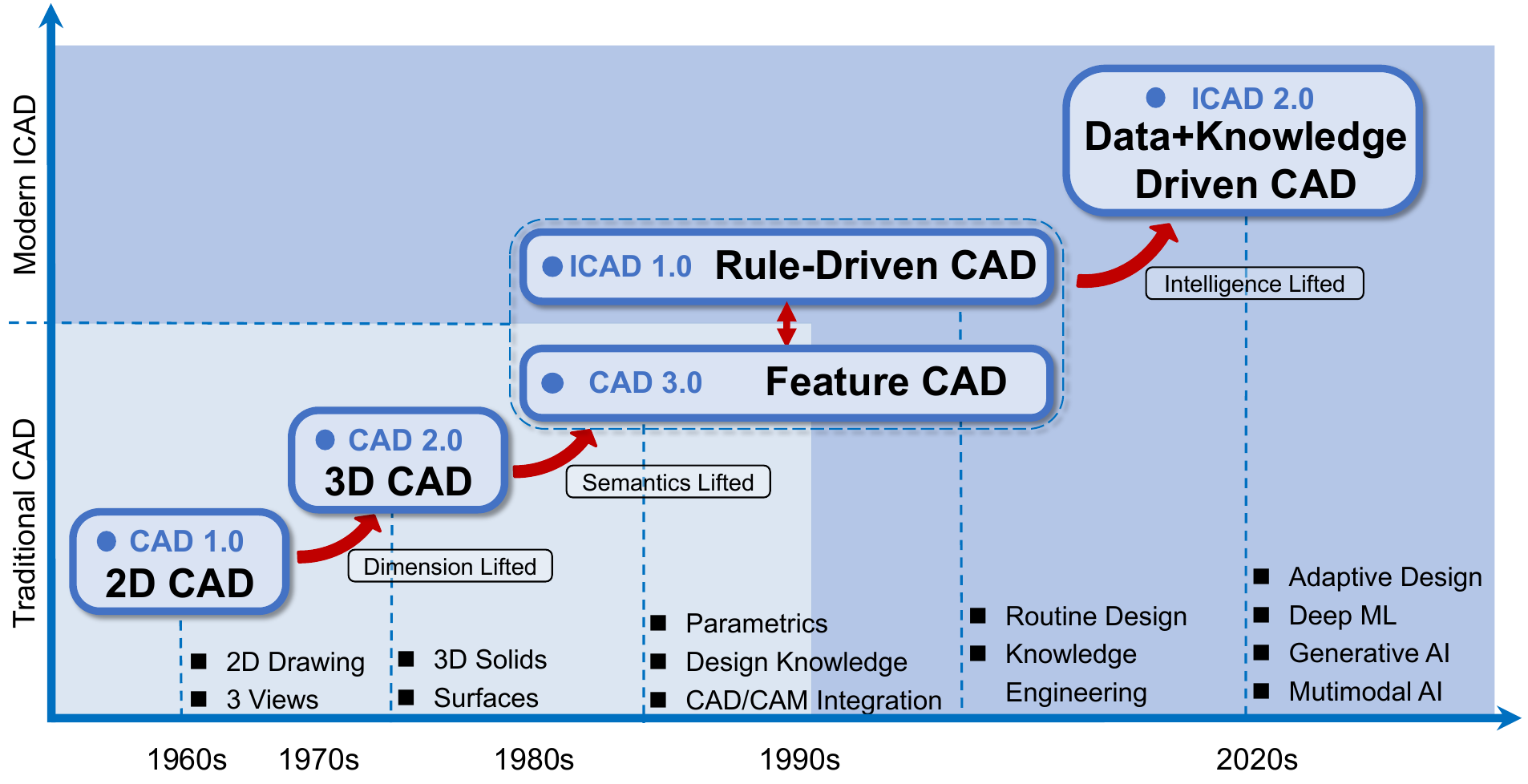}
\caption{Historical development of CAD. Note that certain important stages, such as internet-based and collaborative CAD, are omitted due to their irrelevance to this paper.}
\label{fig:cad-evolution}
\end{figure*}

Over the six decades of CAD history, rapid development was seen in the first 40 years, particularly in geometric modeling, constraint solving, and feature recognition~\cite{zou2023variational}, see Fig.~\ref{fig:cad-evolution} for more details. However, progress has since stagnated, with only refined algorithms and broader applications. This is linked to the initial vision of CAD outlined by Coons~\cite{coons1960computer, coons1963outline}: 
\begin{displayquote} 
``...a system that would in effect join man and machine in an intimate cooperative complex, a combination that would use the creative and imaginative powers of the man and the analytical and computational powers of the machine..." 
\end{displayquote} 
CAD so defined effectively divides the design process into creative and mechanical tasks, with designers overseeing the former and CAD systems managing the latter (see~\cite{coons1960computer} for the reasoning behind this division). While current CAD techniques align well with this vision, further progress requires going beyond it to address the fundamental gap between the unstructured, creative nature of the design process and the rigid, prescriptive nature of computer technology.

Recent advances in artificial intelligence (AI) may bridge this gap and revitalize the intelligent CAD (ICAD) concept from the 1980s and 1990s~\cite{tomiyama2007intelligent,dixon1987expert}. However, the current generation of ``AI+CAD" will differ significantly from its predecessor. There are good reasons for this. 
First, AI techniques have evolved~\cite{haenlein2019brief}. Previous ICAD relied on knowledge engineering, which could only present designs prescribed in the hand-crafted knowledge base. In contrast, current ICAD utilizes data-driven machine learning to extract design patterns and generate innovative new ones. 
Second, the engineering landscape has shifted~\cite{fogliatto2012mass}. In the 1980s and 1990s, mass production dominated, emphasizing standardized products and routine design. Today, the focus shifts toward custom production, prioritizing personalization and innovation. As a result, the previous vision of ICAD is found insufficient, necessitating a new ICAD framework with fundamentally different capabilities, implementations, and applicability.
In what follows, we will refer to the previous generation as ICAD 1.0 and the next generation as ICAD 2.0.

\begin{figure*}[t]
\centering
\includegraphics[width=0.6\textwidth]{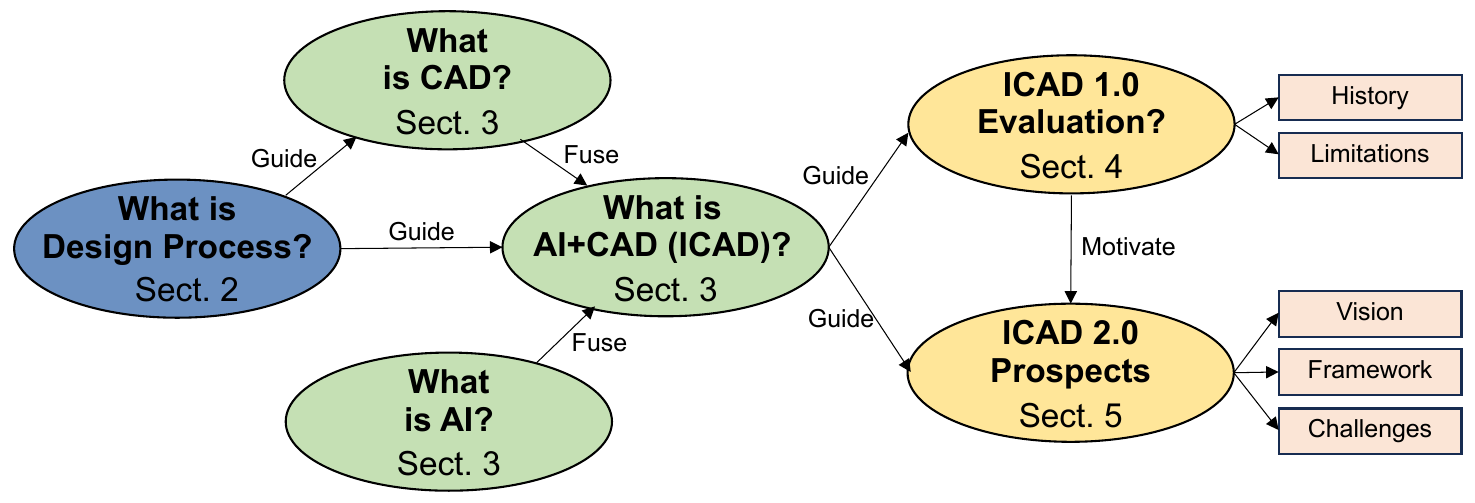}
\caption{Paper framework.}
\label{fig:paper-framework}
\end{figure*}

Despite the growing interest in ``AI+CAD", its future direction remains unclear. What should we anticipate from ICAD 2.0, what strategic pathways should we pursue for its implementation, and what potential challenges might arise? 
This paper seeks to address these questions by examining the connections between the design process, CAD limitations, and AI capabilities. Fig.~\ref{fig:paper-framework} illustrates the paper's framework, which begins with an analysis of the design process. This helps identify design tasks manageable by AI and leads to a definition of ICAD. It then assesses the historical development and limitations of ICAD 1.0, and uses these insights to inform our vision, framework, and challenges of ICAD 2.0.

\section{The Design Process}
\label{sec:design-process}
\begin{figure*}[t!]
\centering
\includegraphics[width=0.8\textwidth]{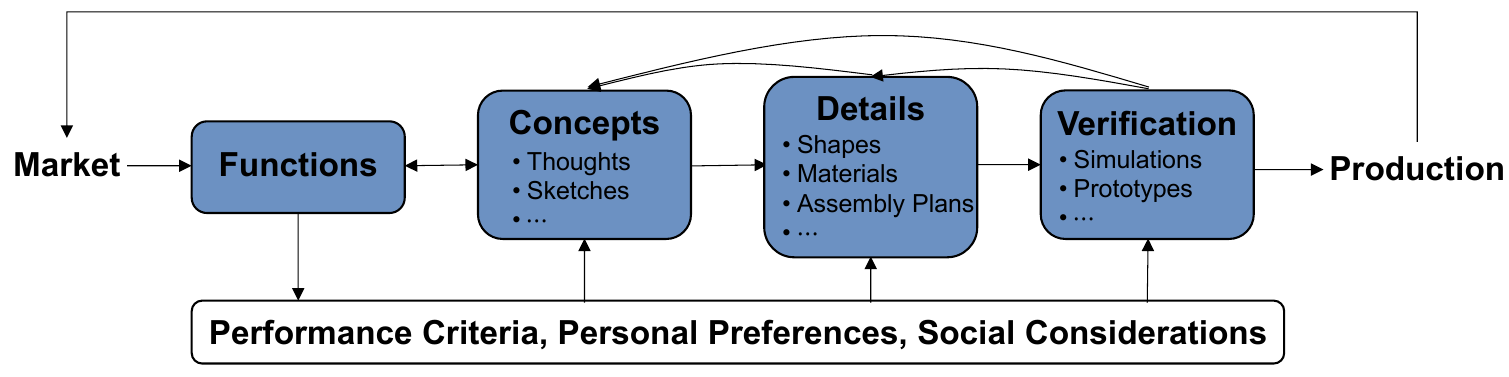}
\caption{An over-simplified illustration of the engineering design process.}
\label{fig:design-process}
\end{figure*}

An in-depth understanding of engineering design and its associated process is crucial for delineating what AI can and cannot accomplish. In a narrow sense, \textit{design refers to the creation of specifications for artifacts and their construction}, ensuring that functional, manufacturability, and aesthetic requirements are met. Designers develop these specifications not randomly but through a structured workflow known as the design process~\cite{pahl1996engineering,cross2021engineering}: in the early phases, they establish the desired functions and initial concepts of the product; later, these concepts are refined to ensure the product meets functional requirements and is manufacturable; and finally, the concepts culminate in comprehensive, concrete representations (e.g., CAD models). Fig.~\ref{fig:design-process} illustrates this process.

Although there is no universally accepted definition of the design process, some common design activities can be identified. For our discussions, the following activities are useful:  functional design (or called requirements analysis), conceptual design, preliminary design, detailed design, and design verification~\cite{shah1995parametric}:
\begin{itemize}
    \item \textit{Functional Design}: Defining the intended use of a design, such as performance criteria.
    \item \textit{Conceptual Design}: Generating initial concepts to address functional requirements.
    \item \textit{Preliminary Design}: Refining concepts into a more solid form, focusing on the overall aspects of the product.
    \item \textit{Detailed Design}: Further refining the concepts to a level where the product's shape is precise, and specifying tolerances, materials, assembly plans, and manufacturing process are possible. This phase produces detailed CAD models and documentation necessary for production.
    \item \textit{Design Verification}: Ensuring that the detailed design meets all functional requirements and manufacturing constraints, such as strength, heat transfer, and vibration, through numerical or physical methods.
\end{itemize}

Using mechanical design as an example, the design process begins with identifying market demands and generating initial design concepts. These concepts are typically expressed through sketches, augmented with symbols, voices, gestures, or the like. This implies that, at the onset of the design, the designer conceives his/her idea not in the precision of mathematics, nor well-crafted statements, but rather in the vagueness of characteristic features, focusing only on high-level semantics.

From the sketches, design analysis can be carried out and accordingly, decisions to keep, to modify, or to discard part or all of the original concept are made. In the early stages, these analytical processes tend to be simplified and qualitative, reflecting the inherent ambiguity of design concepts. At this point, neither the mathematical modeling nor the actual calculations are carried out in great detail; rather, the focus is on drawing conclusions from the models and calculations to provide qualitative guidance for refinement, even if such conclusions are mostly not accurate. This reflects the designer's need to explore alternatives and the inherent uncertainty in some aspects of the design. 

The refined concept enables more precise analysis and decision-making, leading to further iterations of refinement and increasingly accurate analysis.  As this goes on, the designer becomes more certain about the characteristic features of the design, the concept gradually takes a solid shape, and the sketch has more details. This indicates a transition from conceptual design to preliminary design. It is important to note that there is no clear boundary between these two phases; basically, preliminary design involves more detailed modeling and analysis than concept design.

The iterative process culminates in a well-defined design that enables precise mathematical modeling and numerical calculations. At this stage, the sketch becomes settled, with comprehensive details and precision. The designer finalizes geometric specifications while also considering material selection, tolerances, and manufacturing processes. This marks a transition from preliminary design to detailed design, which solidifies the concepts from earlier phases and moves toward a final product that is both optimized for its intended use and viable for manufacturing. The outcomes of the detailed design stage include detailed CAD models, part lists, bills of materials, manufacturing processes, and more.

At the conclusion of the design process are simulations and/or prototypes ensuring that the design meets all functional requirements and manufacturing constraints. Also, continuous design improvement based on post-launch user feedback and real-world performance is important for the design's future iterations, ultimately ensuring the product remains competitive and meets evolving market demands.

Typically, the above design activities---from concept, through analysis, evaluation of the analysis, decision to modify the concept, and finally to a more developed concept---form loops that are traversed again and again, until eventually the design satisfies functional requirements and assumes a well-defined form, see Fig.~\ref{fig:design-process}. Also, the design moves forward at a mixed level of detail. Critical parts may be designed long before less important ones. Some aspects of detailed design may be postponed to production engineering, where the manufacturing constraints may be better taken into account. In a word, the design activities are never sequential, and the design process is a dynamic interplay between creativity and technical rigor.

Summarizing the discussions, we can have the following observations that are useful for studying ``AI+CAD": 
\begin{itemize}
    \item \textit{Ambiguous and Uncertain}: Early design phases often involve vague, incomplete, or even inconsistent concepts.
    \item \textit{Trial and Error}: The design process consists of repeated cycles of refinement and evaluation for experimenting with alternatives until a viable/optimized option is reached.
    \item \textit{Creative Yet Mechanical}: Designers engage in highly creative tasks during the initial stages, but as the design progresses, they transition to more mechanical tasks, such as CAD modeling and finite element analysis.
\end{itemize}

\section{Intelligent CAD in What Sense?}
\label{sec:icad-definition}
Throughout the design process, some tasks are well-suited to computers and AI, while others require human intervention. To draw a distinction between them, we first define CAD and AI individually and then explore their synergistic potential. We emphasize how modern AI allows us to expand CAD capabilities beyond mechanical tasks to include creative tasks in the design process. (It is important to note that achieving universally accepted definitions of CAD, AI, and ICAD is nearly impossible. Thus, the definitions provided here are tailored for our discussions.)

\subsection{What is CAD?}
\label{sec:cad-definition}
In the literature, CAD is consistently defined in a very abstract manner, e.g., the use of computers to aid the design process~\cite{lee1999principles}. While such high-level definitions are absolutely correct, they lack the essential insights for grasping and implementing CAD. Some improvements are, for example, the use of computers to aid the creation, analysis, or optimization of a design. While seemingly more specific, such definitions fall short as they merely enumerate certain tasks involved in the design process without providing useful understanding. Some even define CAD in terms of metaphors like ``Designer Assistant", which can lead to further confusion~\cite{maccallum1990does}.

\begin{figure*}[t]
\centering
\includegraphics[width=0.6\textwidth]{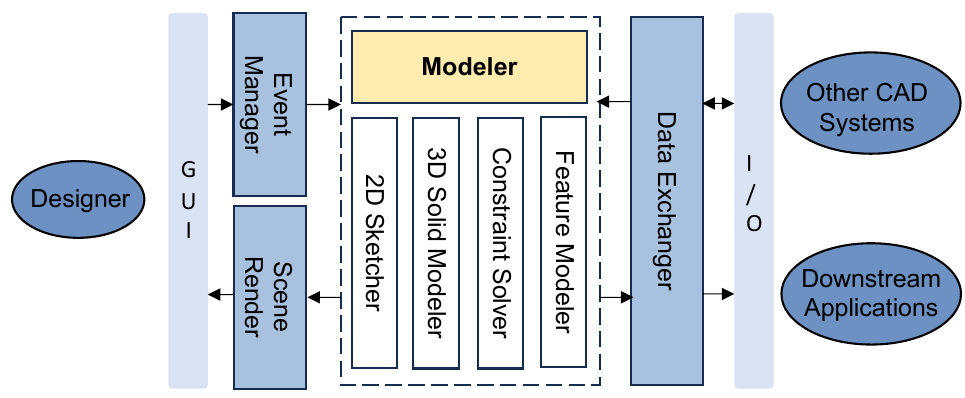}
\caption{Schematic CAD software framework.}
\label{fig:cad-software-architecture}
\end{figure*}

This work opts to approach this problem from the perspective of what fundamental functions CAD must have. Two such functions have been identified: (1) design representations, which capture and embody the design, and (2) their manipulations, which articulate the design process. Steve Coons, in his seminal vision of CAD, described CAD as a designer-computer cooperation complex~\cite{coons1963outline}. For such cooperation to occur, a shared design representation that is understandable to both the designer and the computer is a prerequisite. Thus, design representations are the most essential element of CAD. 

Equally crucial are the manipulations of these representations, as the progression of a design from conceptual to concrete relies on how these representations are manipulated, as well as the decision-making behind it. These manipulations and decisions manifest as activities such as modification, analysis, optimization, and planning within the design process. Consequently, the notions of representations, manipulations, analysis, optimization, or the like are interconnected and build upon each other in a hierarchical manner: design representations and manipulations form the foundation, while the other tasks are advanced functions supported by this base.

Following straightforwardly, CAD can be defined as the use of computers to aid in the creation, analysis, and modification of design representations, including the decision-making associated with them. From an engineering and computational point of view, this can be further elaborated as using computers to aid the construction, analysis, and editing of product models. A product is a physical artifact, while a product model is its digital counterpart~\cite{qiang2019variational}. As such, CAD focuses on modeling all aspects of product information necessary to support the entire design process, from initial concepts through to final production. Here we use a more inclusive definition of modeling to encompass the creation and edits of product models, as well as the associated decisions.
In this context, the product model serves as the common ground for collaboration between the designer and the computer, while the modeling process embodies the design process.

\begin{figure}[b!]
\centering
\includegraphics[width=0.48\textwidth]{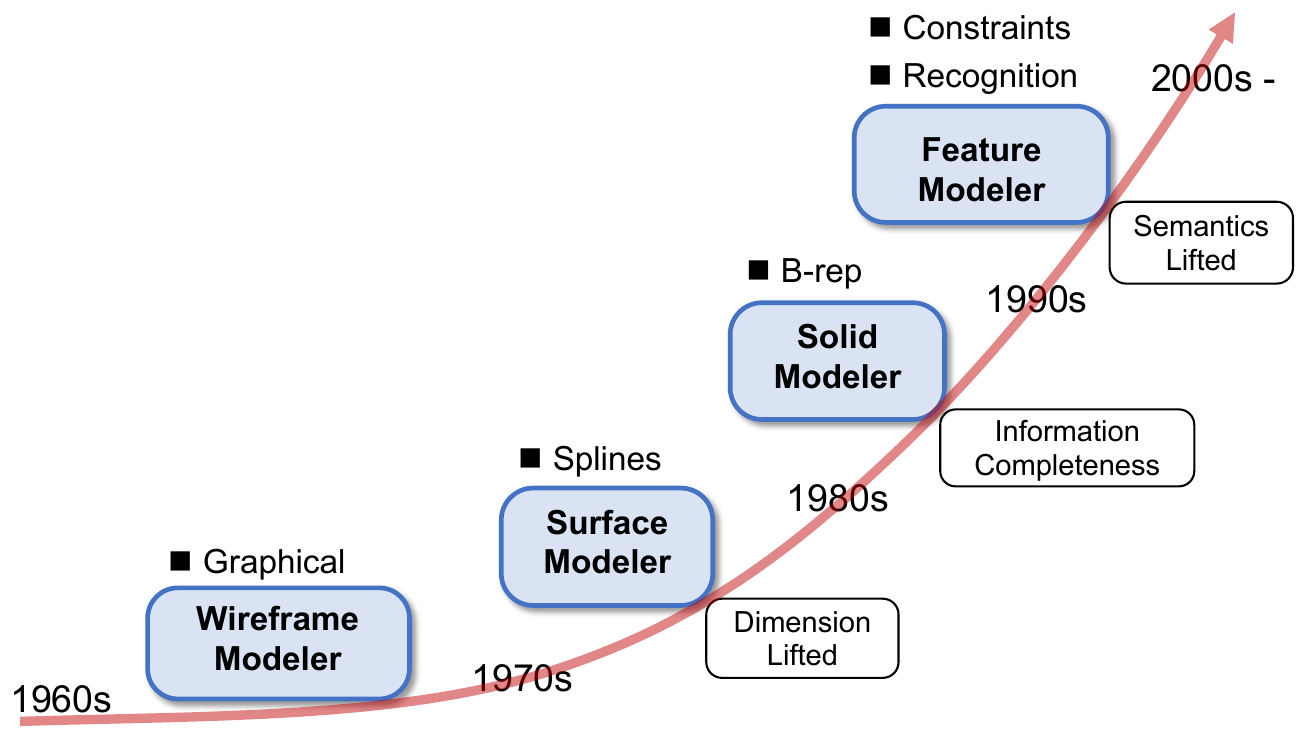}
\caption{Historical development of CAD modeler.}
\label{fig:cad-modeler-development}
\end{figure}

This representation-centric definition also clarifies why advancements in CAD techniques are primarily characterized by improvements in CAD modelers~\cite{shah1995parametric,piegl2005ten,zou2023variational}, meaning that many consider modeling more fundamental over other functionalities in CAD software. Fig.~\ref{fig:cad-software-architecture} shows an architecture of modern CAD software, highlighting the central role of the modeler module. Fig.~\ref{fig:cad-modeler-development} illustrates the historical development of CAD modeling techniques, from wireframe modeling to surface modeling, solid modeling, and feature modeling, reflecting the broader evolution of CAD as recognized by many researchers~\cite{shah2001discourse,farin2014curves,zou2019push}.

This definition also highlights the current limitations of CAD systems. Contemporary design representation schemes, such as solid and feature models, are sufficient to describe detailed designs, especially geometric complexities. However, they struggle with vague, incomplete product information like conceptual ideas. Digital representations/models for conceptual ideas remain underdeveloped for two primary reasons: (1) concepts inherently lack structure and clarity, posing difficulties for computer representation and processing; and (2) the initial vision of CAD, articulated by Coons in the early 1960s~\cite{coons1963outline}, emphasized mechanical and repetitive tasks in the design process, particularly engineering drawings, FEM analysis, and NC programming, which only become relevant after the detailed design phrase.

\subsection{What is AI?}
\label{sec:ai-definition}
AI has been defined from various perspectives, including problem-solving, learning, decision-making, and more. In our context, AI is to be understood as a fusion of ``artificial" and ``intelligence". The term ``artificial" simply means that it is man-made rather than natural, while ``intelligence" refers to the capability of perceiving information, retaining it as knowledge, and applying it adaptively in new environments~\cite{deary2020intelligence}. Essentially, intelligence involves the acquisition and application of knowledge.

\begin{figure}[htbp]
\centering
\includegraphics[width=0.48\textwidth]{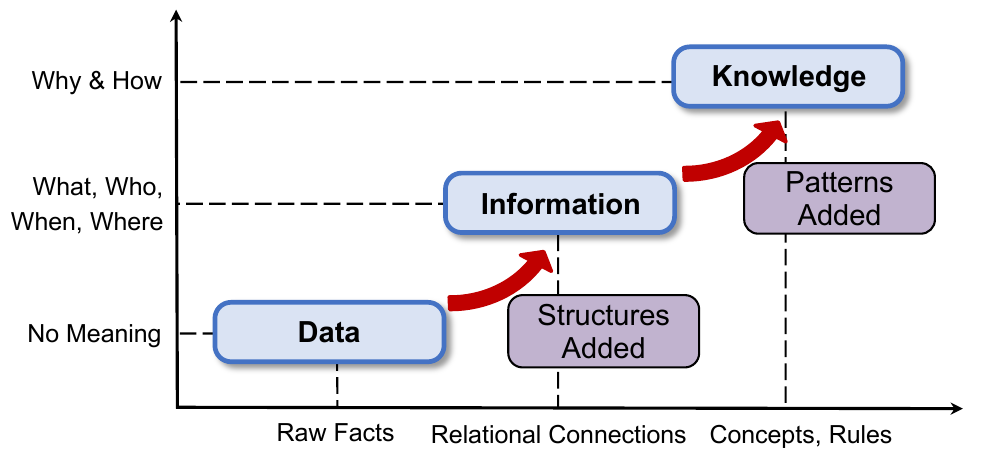}
\caption{Illustration of the data-information-knowledge hierarchy.}
\label{fig:knowledge-hierarchy}
\end{figure}

Knowledge, again, is an abstract term without a universally accepted definition. Understood in the context of ``AI+CAD", we may grasp it from the hierarchy of data-information-knowledge shown in Fig.~\ref{fig:knowledge-hierarchy}~\cite{bellinger2004data}:
\begin{itemize}
    \item \textit{Data}: Raw facts such as numbers, words, or images. Data has no meaning on its own and requires interpretation to become useful.
    \item \textit{Information}: Structured data to give meanings by way of relational connections. It can answer questions like ``what," ``when," "where," and ``who". A relational database is a typical example.
    \item \textit{Knowledge}: Patterned information to give understanding by way of concepts (or abstract representations), relationships, and even rules. It identifies recurring structures within information, generalizes them to abstract concepts, reveals their interactions, and formalizes the underlying rules. By leveraging patterned information, recognition, prediction, decision-making, and problem-solving can be done. As such, knowledge can answer questions like ``why" and ``how".
\end{itemize}

To acquire and apply knowledge, diverse AI approaches have been developed, see~\cite{nilsson2014principles} for their details. Among them, symbolism AI and connectionism AI are more relevant to CAD. Symbolic AI relies on the hand-crafted knowledge base and reasoning rules, with expert systems being a typical example~\cite{jackson1990introduction}. These systems are well-suited for prescriptive reasoning and routine design. In contrast, connectionism AI utilizes interconnected networks of artificial neurons to learn patterns from data. Deep learning, a prominent example, has revolutionized tasks such as image classification, generation, and natural language processing~\cite{lecun2015deep}. 
Increasingly, these approaches are not used individually but are integrated to leverage their combined strengths and optimize outcomes across various applications.

\subsection{What is AI+CAD?}
\label{sec:icad-definition}
Having established a representation-centric definition of CAD and a knowledge-based understanding of AI, we are now ready to explore the concept of ``AI+CAD". This concept revolves around the synergistic relationship between ``aid" and ``intelligence". ``Aid" involves the processes of constructing, analyzing, and editing product models, including both the actions performed and the decisions behind. ``Intelligence", on the other hand, focuses on leveraging knowledge. Thus, ``AI+CAD" can be defined as the acquisition of design knowledge and its application to facilitate the construction, analysis, and editing of product models. However, like all previous definitions of ICAD in the literature~\cite{tomiyama2007intelligent}, this definition remains somewhat abstract and lacks practical insights into its essential nature.

To deepen our understanding, we must investigate what CAD needs and what AI can contribute. As previously discussed, CAD is fundamentally concerned with the modeling of a product's information to support the entire design process. Therefore, it is essential to address the following key questions to assess CAD's needs:

\textit{Question \#1 (Informational Completeness)}: Can current CAD models (i.e., design representations) automatically answer queries about the product's properties from any tasks in the design process?

\textit{Question \#2 (Tool Usability)}: Are existing CAD modeling tools (i.e., design manipulation) effective and user-friendly for creating, analyzing, and modifying CAD models?

Addressing the first question, a product's specifications in engineering design primarily encompass its spatial information, augmented by secondary physical details such as material, hardness, and manufacturing processes~\cite{jayanti2006developing}. Some might argue that cognitive information---how the product is mentally conceived by designers---should also be considered. However, cognitive information still pertains to the geometric and/or physical aspects of the product, albeit in a vague and incomplete form. It is also worth noting that while management information is relevant to design, it relates more to product lifecycle management than to CAD itself. Therefore, it is not included in our discussion.

To represent spatial information, solid models~\footnote{A solid is a subset of $\mathbb{R}^3$ that is bounded, regular, semi-analytic, and has a manifold boundary~\cite{requicha1983solid}. A solid model is a computerized representation of a solid. The mainstream schemes for solid modeling are boundary representation, which describes a solid using its boundary with non-solid space, and constructive solid geometry, which represents a solid through successive combinations of primitive geometries. Refer to~\cite{qiang2019variational} for a thorough introduction to solids.} have been developed. Once successfully constructed, solid models can answer any geometric queries such as point membership classification and inertia calculation~\cite{shapiro2002solid}. However, constructing solid models requires comprehensive geometric details and precision, which limits their utility until the detailed design stage. Before this stage, the unsettled geometric details hinder successful solid model construction. Thus, current CAD models provide only partial completeness in spatial information: they are insufficient before the detailed design phase but become fully informative thereafter.

To represent physical information, feature models~\footnote{A feature is a recurring, engineering-meaningful geometric portion of a product~\cite{shah2001discourse}. It groups geometric entities within a solid model, serving as a container for storing physical information. Some researchers prefer the term ``design intents" to describe the intended use or manufacturing processes of the product, which are inherently physical.} have been developed. Using features, physical information can be effectively integrated with geometric information~\cite{zou2023variational}. However, current feature models are prescriptive, limiting queries to hand-crafted rules. Moreover, since features are based on solid models, they are useful only after the detailed design stage. Consequently, current feature models are also partially complete in physical information: insufficient before detailed design and only prescriptively informative thereafter. 

With informational incompleteness comes the inability to effectively manipulate geometric and physical design information during the early stages of design, which ties to the second question. Even in the detailed design phase and beyond, where solid and feature models are assumed to be beneficial, their manipulation remains complex~\cite{zou2019push,shah2001discourse}. Particularly problematic is the lack of decision support tools to assist in determining whether to keep, remove, or modify parts of solid and feature models, and this is due to the absence of automatic analysis tools and intelligent interpretation capabilities. Additionally, manipulating solid models is often non-intuitive and inefficient. Feature model manipulation, such as feature recognition and modification, is limited by predefined rules. Given these limitations and the nature of the design process summarized at the end of Sect.~\ref{sec:design-process}, current CAD techniques are more effective as design documentation tools rather than as true design support systems.

To address the issue of informational incompleteness, next-generation CAD models should be (1) dynamic in geometric information and (2) flexible in physical information. Dynamism means that geometric models evolve as the design process progresses, transitioning from initially vague concepts (which cannot be captured by precise geometries), to detailed geometries such as curves and surfaces, and ultimately to machine instructions where shapes are derived from machine movements rather than being explicitly defined. Hence, CAD models need to extend beyond solid models to include both vague concepts and machine instructions. Flexibility refers to the ability to exceed the prescriptive constraints of current feature models, enabling unrestricted physical queries throughout the design process. 

\begin{figure*}[t!]
\centering
\includegraphics[width=0.95\textwidth]{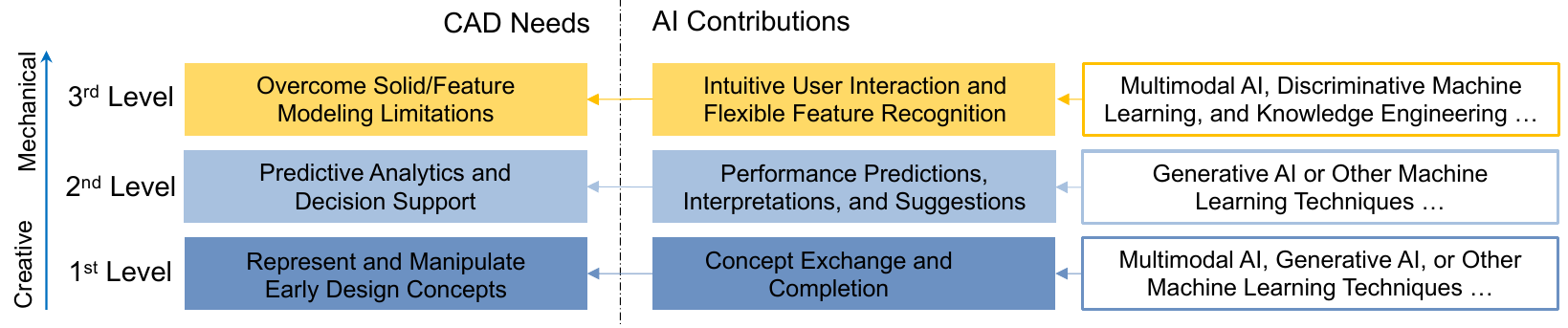}
\caption{The three-level structure of CAD needs and AI contributions.}
\label{fig:cad-needs}
\end{figure*}

To address the issue of tool inefficacy, next-generation CAD modelers should (1) support early design phases by accommodating vague concepts and facilitating their continuous, iterative exchange between the designer and computer; (2) provide intelligent decision support through effective integration of predictive analytics and their contextual interpretations; (3) offer intuitive solid modeling user interactions to ensure users can create and modify geometric models with minimal training and effort; and (4) allow flexibility in feature modeling to enable a more adaptive and responsive approach to non-routine or creative design.

Summarizing the CAD needs outlined above reveals a three-level structure, as depicted in Fig.~\ref{fig:cad-needs}. At the first level is the representation and manipulation of designs in their early stages, a topic where current CAD techniques are notably inadequate. At the opposite end, the third level addresses the limitations of current solid and feature modeling techniques, emphasizing the need for more intuitive and flexible modeling tools, an area with a solid foundation for improvement. In between these extremes, the second level involves decision support for design manipulation, where CAD tools not only perform routine functions such as design analysis but also engage in more creative tasks, such as interpreting analysis results and making informed recommendations.

With the CAD needs in place, we can now explore how AI can contribute and what ICAD is. At the core of the first-level CAD needs is the iterative exchange of concepts and decisions between the designer and the computer. The computer should be capable of accepting all forms of input from the designer, whether verbal, gestural, graphical, symbolic, or numerical. It should also be able to provide meaningful feedback for refinement or completion. As highlighted at the end of Sect.~\ref{sec:design-process}, these inputs are often vague, incomplete, or even inconsistent. AI, as described in Sect.~\ref{sec:ai-definition}, is concerned with acquiring and applying knowledge, making it adept at recognizing and handling abstract patterns rather than precise details, at least in principle. This positions AI well to enhance concept exchange and completion tasks, although current AI technology may not be able to fully realize this potential.

However, AI's role in decision-making during the iterative concept exchange and completion process remains somewhat limited. In engineering design, these decisions often involve personal or social considerations, such as value judgments and preferences, which extend beyond mere data-driven analysis or numerical optimization. These elements typically require human insights and subjective evaluation, which AI cannot easily replicate. (The authors personally believe that computers should not, and perhaps never will, possess such capabilities, as they reflect wisdom---a quality we hope remains uniquely with humans.)

The second-level CAD needs are closely linked to the ``Trial and Error" highlighted at the end of Section~\ref{sec:design-process}. This process involves iterative cycles of evaluation and refinement, experimenting with various solutions until a viable or optimized one is found. Key tasks at this level include predicting design performance, interpreting the predictions, and providing recommendations for refinement. Together with tasks at the first level, these efforts effectively streamline the iterative process across (1) design concepts and decisions, (2) predictive analytics, and (3) refinement recommendations. AI can significantly enhance all these tasks. For predictive analytics, modern AI techniques can improve traditional design analysis tools (e.g.,~\cite{zou2023xvoxel}) with more efficient quantitative analytics. They also enable the recognition of structural elements and their association with qualitative analytics, even when the design is not yet fully developed into precise models. For refinement recommendations, generative AI excels in creating design alternatives and proposing refinements that align with both quantitative data and qualitative insights.

For third-level CAD needs, AI offers immediate and substantial assistance. It can enhance interactions between designers and CAD models through natural and intuitive interfaces. For instance, AI can facilitate seamless communication using speech recognition and advanced graphical interfaces. In feature modeling, AI can go beyond traditional prescriptive features by enabling flexible feature definition and recognition. This capability allows AI to adapt to varying design requirements and interpret a wider range of features that may not be predefined or explicitly specified.


Based on the preceding discussions, we arrive at a definition of ICAD as: \textit{a collaborative system that integrates the designer's creative and evaluative abilities with the computer’s recognition, reasoning, and generative capabilities---derived from its programmed and/or learned knowledge---to aid in the construction, analysis, and editing of product models.} 

ICAD should facilitate continuous and iterative exchanges of vague concepts, predictive analytics, and refinement recommendations concerning the design's geometric and physical aspects; importantly, the exchanged information is allowed to be incomplete, inaccurate, or even inconsistent. As such, we are creating tools for the designer from the intensional perspective of how designers think rather than for the mathematics involved. When effectively implemented, such a system enables a human-computer collaboration that offers a level of freedom and precision far exceeding that of traditional CAD systems.

What distinguishes the ICAD here from traditional CAD (i.e., the first CAD vision by Coons~\cite{coons1963outline}) is the added creativity to the computer, moving beyond mere repetitive mechanical tasks. Compared to previous generation ICAD (i.e., the second CAD vision by Tomiyama~\cite{tomiyama2007intelligent}), which also sought to leverage knowledge and include creative tasks, this new ICAD vision extends beyond the extensional perspective into the intensional realm of design. Previously, ICAD systems focused on observing what designers do and capturing recurring aspects of their work in the knowledge base and reasoning rules. This, however, leaves out what designers really think and do. In contrast, the ICAD here emphasizes seamless exchanges of concepts and decisions, enabling a richer sharing of the intensional aspects of design. 

\section{Historical Development: Intelligent CAD 1.0}
\label{sec:icad1.0}
\begin{figure*}[t]
\centering
\includegraphics[width=0.7\textwidth]{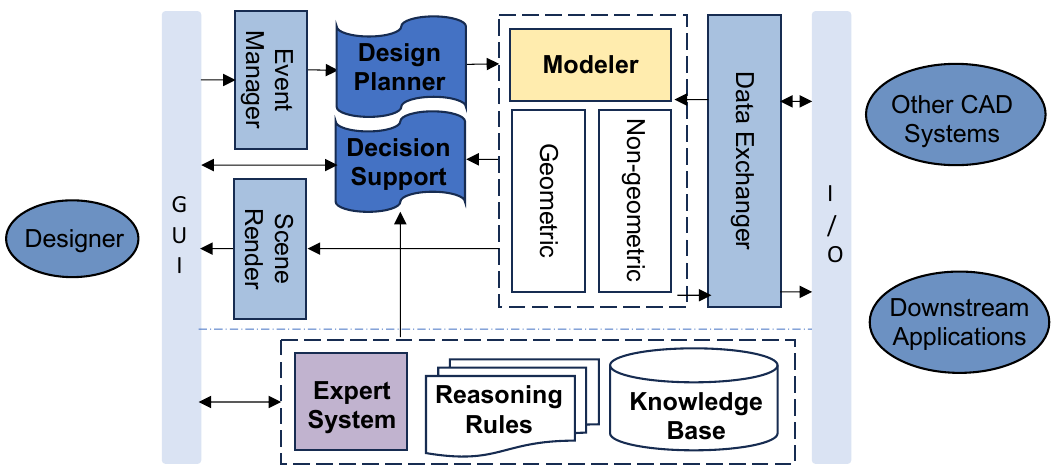}
\caption{Schematic diagram of ICAD 1.0 framework.}
\label{fig:icad.10-framwork}
\end{figure*}

Historically, ICAD was understood and implemented from the perspective of knowledge engineering due to its prominence in AI in the 1980s and 1990s~\cite{tomiyama2007intelligent,maccallum1990does,nilsson2014principles}. At that time, the terms ICAD, knowledge-based CAD, and expert CAD were often used interchangeably. Knowledge engineering simulates humans' expertise and problem-solving abilities utilizing knowledge bases and reasoning rules~\cite{jackson1990introduction}. Integrating it into CAD by means of domain knowledge shells enabled the automation of routine tasks and improved design consistency. Fig.~\ref{fig:icad.10-framwork} shows a typical example of using knowledge shells in ICAD software at that time. There are excellent references available for a comprehensive review and summary of ICAD 1.0, such as~\cite{la2012knowledge,tomiyama2007intelligent,ten2012intelligent}. Therefore, we provide here only a brief overview of its development history and focus primarily on its limitations.

\begin{figure}[htbp]
\centering
\includegraphics[width=0.48\textwidth]{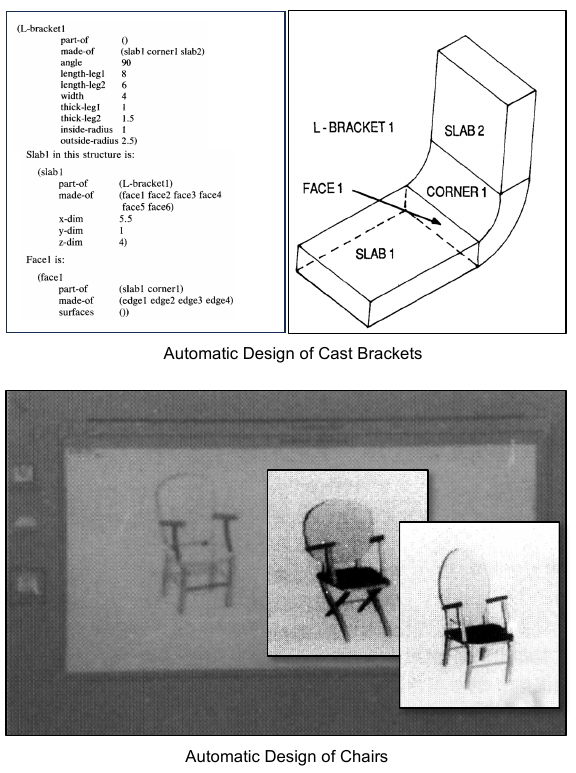}
\caption{Examples of applications of expert CAD systems: (top) the bracket design example from Dixon et al.~\cite{dixon1987expertG}; (bottom) the chair design example from Pan et al.~\cite{pan1996icad}}
\label{fig:expert-cad-example}
\end{figure}

In the 1980s, the application of knowledge engineering to CAD became a popular research topic in engineering, inspired by the successes of expert systems such as MYCIN (for medical diagnosis) and DENDRAL (for chemical analysis). This period also saw the emergence of the concept of ICAD, with Tomiyama~\cite{tomiyama2007intelligent} possibly being the first to formally coin this term in 1983. He described ICAD as the use of knowledge engineering to automate certain tasks within the design process. Several practical ICAD systems have also been developed, often highly specialized and tailored to routine design tasks. For instance, Dixon et al.~\cite{dixon1984expert} created a knowledge shell for V-belt design. Pan et al.~\cite{pan1996icad} designed an expert system for chair design, see Fig.~\ref{fig:expert-cad-example}. Many of these systems found practical applications, particularly in parametric design and constraint management tasks.

In the 1990s, research efforts shifted towards developing domain-independent knowledge bases and model-based reasoning techniques~\cite{tomiyama2007intelligent,la2012knowledge}. However, researchers soon realized that creating an exhaustive knowledge base and rule set for complex systems like CAD, particularly for creative design problems, was impractical. Expert systems proved effective primarily in well-defined domains, such as routine assembly or variant design based on predefined configurations (e.g., constraint-based design~\cite{zou2020decision}). The development of the feature-based CAD paradigm can be seen as a partial outcome of these research efforts~\cite{rosen1993feature}.

In the late 1990s and thereafter, research on ICAD 1.0 faced stagnation due to several inherent challenges and limitations~\cite{shah1995parametric,tomiyama2007intelligent}:
\begin{itemize}
    \item \textit{Narrow Domain Knowledge and Rigid Rules}: ICAD 1.0 systems struggled with creating comprehensive knowledge bases and flexible reasoning rules. They could only present design solutions prescribed by the knowledge base and rules, making them suitable for domain-specific, routine design problems but not for creative or dynamic ones.
    \item \textit{Shallow-Level Knowledge}: ICAD 1.0 systems relied on symbolic reasoning and were not adept at handling 3D spatial information. Specialized rules were needed to address these limitations, which made the overall system brittle and difficult to maintain.
    \item \textit{Limited Performance}: Even for design problems within their designated domain,  ICAD 1.0 systems could not deliver genuinely expert performance. They approached design knowledge extensionally, without delving into the intensional realm of design and understanding what designers truly think and do.
    \item \textit{Hard Knowledge Maintenance}: Product design evolves rapidly. Keeping the CAD knowledge base up-to-date and relevant was challenging, often requiring substantial manual effort and sometimes even being nearly impossible.
    \item \textit{Unintuitive User Interaction}: ICAD 1.0 systems were often not user-friendly, making it difficult for non-experts to fully leverage the intelligent features.
\end{itemize}

Despite these challenges, the foundational work in expert CAD during the 1980s and 1990s laid the groundwork for subsequent understanding of ICAD. These limitations also underscored the need for advancements beyond early ICAD vision, attaining more sophisticated approaches that integrate modern AI and CAD techniques and deliver genuine design intelligence to enhance the design process.

\section{Prospective Development: Intelligent CAD 2.0}
\label{sec:icad2.0}
Based on the definition of ICAD and the limitations of ICAD 1.0, we envision a new-generation system where a designer is seated at a computer, sketching the proposed design with augmentations such as speech, gestures, symbols, and numbers. The computer interprets and reasons about these potentially inaccurate, incomplete, or inconsistent inputs, and performs (not necessarily accurate) analyses such as strength, clearances, and manufacturability. It then generates recommendations to refine or auto-complete the design. Based on these recommendations, the designer evaluates their alignment with intended functions, aesthetic preferences, and social considerations, makes decisions on design refinements, and communicates these decisions to the computer for another round of concept understanding, predictive analytics, and refinement recommendations. This iterative process allows the design to progress at an exponential rate, with the designer needing only to make decisions at will.

As the design approaches its final form, the designer's vision becomes more defined, and the computer’s analyses, interpretations, and recommendations become increasingly precise. Fewer alternatives are needed, shifting the focus from creative exploration to mechanical computation. Ultimately, this collaborative effort results in a highly refined solid model that integrates technical accuracy with the designer’s intention. At this point, the computer begins to operate autonomously, first recognizing features from the solid model, and then using these features to lift from the geometric level to the physical (or semantic) level for design verification and process planning. 

Additionally, we anticipate that the computer will learn from the designer’s preferences through ongoing interactions. By analyzing behavioral patterns and integrating feedback, the computer can customize the design process to individual needs, adapting its tools and recommendations to align with the designer’s style. This personalized approach ensures that the CAD system evolves with the designer, providing a more intuitive and seamless design collaboration experience.

The design collaboration outlined above does not yet exist in current CAD systems and remains largely within the realm of vision. However, rapid advancements in AI technology are steadily making this vision more achievable. With deep learning, multimodal AI, and generative AI, a clear path toward achieving this goal is emerging. The following provides a potential framework for turning this vision into reality.


\subsection{A Conceptual Framework}
\label{sec:icad2.0-conceptual framework}
\begin{figure*}[t]
\centering
\includegraphics[width=0.98\textwidth]{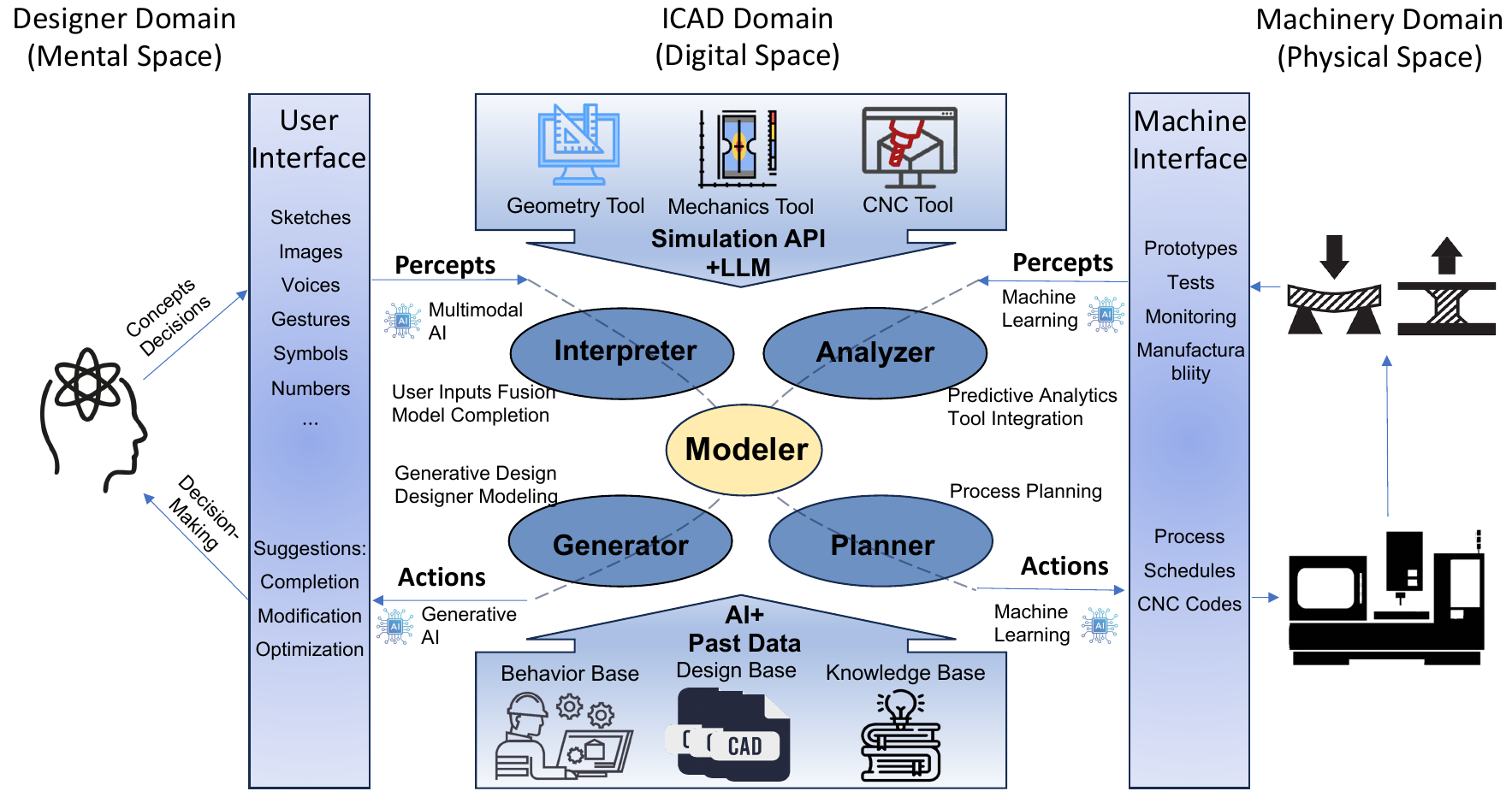}
\caption{A concept framework of ICAD 2.0.}
\label{fig:icad2.0-framework}
\end{figure*}

Fig.~\ref{fig:icad2.0-framework} presents the proposed ICAD framework, which consists of five core modules and two supporting modules. The core modules work in a closed-loop manner, connecting the designer, ICAD, and the production tool. The interpreter module captures and understands the designer's inputs, while the modeler module converts these understands into geometric and physical models to be stored in the computer for further process. The analyzer module then assesses the design's performance based on these models, and the generator module gathers current states and evaluations to produce refinement recommendations. Finally, the planner module translates these models into process schedules and machine instructions, directing the machine tools to create the final products or prototypes. There are two additional supporting modules. The tool library module facilitates integration with other engineering tools and platforms, especially for simulation tools. Meanwhile, the past data module helps ICAD to learn from historical designs and user interactions to provide personalized and adaptive recommendations.

\textbf{The Interpreter Module.} It accepts various forms of input from the designer, including graphical, verbal, gestures, symbolic, and numerical data. It then synthesizes these inputs to form a coherent understanding of the design’s intent and refines vague and incomplete concepts into preliminary geometric/physical models that encapsulate the designer’s vision. Additionally, it may autocomplete concepts to ensure a more appropriate representation of the design for tasks like engineering analysis.

Multimodal AI, leveraging advanced natural language processing (NLP) and computer vision (CV) techniques, forms an effective foundation for the above functions. NLP facilitates the accurate interpretation of verbal and symbolic inputs, while CV handles graphical and gesture-based ones. Additionally, machine learning algorithms can enhance the synthesis process by learning from previous design interactions and predictions, thereby improving the module’s ability to autocomplete concepts and refine preliminary models. These models not only reflect the designer’s vision but also provide a solid foundation for further refinement and engineering analysis.

\textbf{The Modeler Module.} This module is responsible for storing, creating, and manipulating geometric and physical models of the design. These models must be dynamic, i.e., accommodating vague and incomplete models generated during the early design stages, as well as precise and comprehensive models developed during and after the detailed design phases. This module should also offer intuitive operations for model construction and manipulation. Additionally, an effective translation mechanism is also needed to convert preliminary models into their more precise counterparts to allow predictive analytics (i.e., interfacing with the analyzer module). Likely, an automatic translation from geometric models to physical models is important for interfacing with the planner module.

Generative AI is especially effective in meeting the requirements outlined above. It can complete preliminary designs into precise models and explore numerous variations by sampling from a learned distribution, serving as an efficient translation mechanism. Additionally, large language models (LLMs) enhance user interaction by translating model manipulation commands into model operations and automating API calls. Machine learning algorithms further support this module by recognizing geometric features and elevating them to feature models. The integration of these AI methods enables ICAD to dynamically adapt to evolving design requirements, providing robust support for both conceptual and detailed modeling tasks.

\textbf{The Analyzer Module.} This module functions like traditional analysis tools~\cite{lee1999principles} but is augmented with AI capabilities. It streamlines conventional simulations of geometric properties, mechanical characteristics, machining performance, and more. Additionally, it collects performance feedback from simulations, prototypes, and post-launch operational data to provide a comprehensive assessment of the design’s effectiveness.

LLMs can significantly enhance this module by automating the integration and collaboration of simulation tools. Machine learning algorithms can synthesize both simulated and real data into refinement insights about the design. Furthermore, these algorithms can advance the simulation paradigm through data-driven or physics-informed methods. For instance, physics-informed neural networks can substantially accelerate simulation speed, offering real-time feedback and predictions on design performance across various conditions. Even when these conditions are not fully defined, machine learning-based predictive analytics can still be applied, although their precision may vary.

\textbf{The Generator Module.} 
This is the most characterizing component of the ICAD framework. It learns design patterns from past designs, user preferences, and best practices. By integrating these learned patterns with programmed design knowledge, the module generates refinement recommendations based on design objectives, current geometric/physical models, and predictive analytics. These recommendations may include suggestions for completing concepts, modifying CAD models, optimizing parameters, or exploring alternative design approaches. With them, the designer only needs to decide which options to pursue and ensure that the final design meets technical requirements and user preferences. This can greatly reduce the iterative workload and speed up the overall design process. 

Apparently, generative AI can play a crucial role in implementing this module. Generative models are capable of producing a diverse range of design refinement recommendations by synthesizing learned design patterns. When augmented by knowledge engineering, these models can offer more targeted and contextually relevant recommendations, ensuring that they are both innovative and technically sound. By leveraging these advanced AI techniques, the generator module can provide designers with a richer array of creative and feasible design alternatives, enhance the exploration of the design space, and accelerate the development of optimized, high-performance solutions. It should be noted that some studies use the term ``generative" to refer to design approaches employing shape/topology optimization techniques. These methods focus on mechanical objective minimization rather than intelligence. In this paper, ``generative" refers to generating new designs by intelligent methods which sample from a learned distribution or by complete a design based on an initial starting point.

\textbf{The Planner Module.} This module is activated once the design is nearly finalized, with precise solid and feature models already made available. Its primary role is process planning, encompassing production schedules, assembly plans, machining instructions, and more. Effective algorithms for this module have already been made available (e.g.,~\cite{zou2013iso,zhang2018integrated}) and can be easily integrated into the ICAD framework. AI can further enhance this integration by learning from existing process plans and generating adaptive plans to meet new requirements.

\subsection{Five ICAD Challenges}
\label{sec:icad-challenges}
To fully implement the ICAD framework, several technical challenges must be addressed. Five primary issues have been identified based on their significance within the framework and the authors' experiences, which introduces a degree of subjectivity.

\textbf{Challenge No. 1: Multimodal Design Communication.} Effective exchanges of concepts and decisions between the designer and the computer are crucial for advancing from the current CAD paradigm to the next-generation ICAD system. Key research problems in this area include:
\begin{itemize}
    \item \textit{Data Fusion:} Integrating the designer's inputs of different modalities (e.g., voice commands, gestures, sketches, 3D CAD models) in a way that preserves their individual contributions while creating a cohesive overall representation.
    \item \textit{Design Intent Capture:} Understanding, reasoning, and predicting the context and intent behind the multimodal inputs.
    \item \textit{Personalization:} Accounting for diverse user preferences and interaction styles, including cultural, cognitive, and expertise differences.
\end{itemize}

\textbf{Challenge No. 2: Generative CAD Modeling.} It involves creating precise solid and feature models from ambiguous design descriptions. This is a critical component of the interpreter and modeler modules within the ICAD framework, particularly for implementing the dynamic CAD model and model completion functionalities. Additionally, the generator module is central to providing creative recommendations for design refinement. Key issues that need to be addressed include:
\begin{itemize}
    \item \textit{Predictable Generation:} Providing mechanisms to control the output (i.e., solid/feature models and design refinement recommendations) so that they are predictable and aligned with user needs.
    \item \textit{Quality Control:} Ensuring the quality and validity of generated CAD models, including precise geometry, correct topology, and well-constrained parametrics. For a detailed definition of CAD model validity, refer to Ref.~\cite{mantyla1984note,hoffman2001decomposition,zou2023variational}.
    \item \textit{Vague Inputs:} Handling incomplete or inconsistent inputs to generative AI models.
\end{itemize}

\textbf{Challenge No. 3: Hybrid Design Knowledge.} The synergy of data-driven learning models with rule-based knowledge models can not only enhance creativity in engineering design but also produce more targeted and contextually relevant recommendations, ensuring that the recommendations are both innovative and technically sound. Some relevant problems that should be considered are:
\begin{itemize}
    \item \textit{Representation Fusion:} Balancing the flexibility and inherent uncertainty of data-driven models with the prescriptive and precise nature of rule-based systems to ensure coherence between them.
    \item \textit{Conflict Resolution:} Addressing discrepancies between data-driven insights and rule-based guidelines to ensure consistent and coherent design recommendations.
    \item \textit{Adaptability:} Ensuring the system remains adaptable and flexible as the design evolves, allowing for dynamic adjustment and continuous improvement.
\end{itemize}

\textbf{Challenge No. 4: Shape-Property-Process Correlation.} Effective predictive analytics are essential for streamlining the modules within the ICAD framework, and establishing shape-property-process correlation is crucial for enabling these analytics. This involves aligning geometric models with material properties and manufacturability. The alignment does not need to be precise or exact; rather, approximate analytics are adequate for guiding design refinement. Additionally, this process should not rely on the availability of detailed geometric models; instead, it should be capable of deriving insights from rough geometric models, which is particularly valuable in the early stages of design. Key issues to address include:
\begin{itemize}
    \item \textit{Shape-Property-Process Mapping:} Developing AI-based models to link geometric features with relevant material properties (e.g., strength and stiffness) and manufacturability, which facilitates automated and meaningful analytics.
    \item \textit{Early Design Integration:} Ensuring that predictions and insights are integrated effectively during the early design stages, where detailed geometric models are not yet available.
    \item \textit{Robustness and Dynamics:} Accommodating uncertainties and variations in input data and allowing for flexible adjustments as more detailed information becomes available or as design requirements evolve, ensuring the mapping model remains resilient and responsive throughout the design process.
\end{itemize}

\textbf{Challenge No. 5: Large and High-Quality CAD Datasets.} High-quality and sufficiently large CAD datasets are fundamental for developing AI algorithms to be used by the ICAD framework. Without such datasets, it becomes impossible to train, validate, and refine AI models tailored to engineering design tasks such as learning design patterns, understanding sketches, and making contextually relevant recommendations. Issues in this topic include:
\begin{itemize}
    \item \textit{Data Collection and Accessibility}: Acquiring high-quality, comprehensive, and diverse CAD datasets, particularly 3D models. (Unlike image or text datasets, there is a lack of ``big data" in CAD models. Proprietary data restrictions, confidentiality concerns, and limited access to industry-specific information exacerbate this issue. While few-shot learning offers a potential solution, adapting it effectively to the CAD and engineering design domains remains unclear.)
    \item \textit{Annotation and Quality Control}: Labeling CAD data for effective AI model training and validation. (This demands significant manual effort and domain expertise to correctly annotate geometric features, design constraints, and other attributes. Although unsupervised learning approaches might alleviate some challenges, they do not fully resolve all issues. Moreover, even adapting these methods to the CAD domain remains an ongoing challenge.) 
    \item \textit{Data Standardization and Integration}: Ensuring consistency across various sources, formats, and standards of CAD data, e.g., sketches, solid models, and simulation results. (The diverse nature of CAD data requires robust strategies for standardization and integration to attain data coherence.)
    \item \textit{Scalability and Diversity}: Building sufficiently large and diverse datasets that capture various design patterns, industries, and applications to attain robust AI algorithms to be used in ICAD.
\end{itemize}

\subsection{Comparisons with ICAD 1.0}
\label{sec:icad-difference}

The transition from ICAD 1.0 to ICAD 2.0 signifies a major evolution in the approach to design tasks, enhancing flexibility and fostering collaboration. Table~\ref{tab:icad-differences} highlights the key differences between these two generations of ICAD. The following discussion provides a detailed comparison.

\begin{table*}[ht]
\centering
\resizebox{\textwidth}{!}{
\begin{tabular}{p{3cm}|p{6cm}|p{6cm}}
\hline  \hline
\textbf{Aspect}               & \textbf{ICAD 1.0}                                      & \textbf{ICAD 2.0}                                              \\ \hline
\textbf{Definitions and Conceptual Foundations} & Rule-based design automation using knowledge engineering, grounded in the \textit{extensional} view of design (for definitions of extensional and intensional, refer to the last paragraph of Section~\ref{sec:icad-definition})      & Data and knowledge-driven design collaboration utilizing modern AI techniques such as deep learning and generative AI, based on the \textit{intensional} view of design \\ \hline
\textbf{Core Capabilities}    & Automates routine, domain-specific tasks with rigid rules & Facilitates dynamic, interactive design with real-time feedback and personalized recommendations \\ \hline
\textbf{Key Techniques}       & Utilizes expert systems, knowledge shells, and symbolic reasoning & Employs multimodal AI (NLP, CV), generative AI, and contemporary deep learning techniques \\ \hline
\textbf{Impact on Design Process} & Enhances efficiency and consistency within defined domains & Promotes a flexible, adaptive, and creative design process, particularly for the early design stage \\ \hline \hline
\end{tabular}
}
\caption{Summary of essential differences between ICAD 1.0 and ICAD 2.0}
\label{tab:icad-differences}
\end{table*}

\textbf{Definitions and Conceptual Foundations.}
ICAD 1.0, developed during the 1980s and 1990s, relied on knowledge engineering and rule-based systems to emulate expert problem-solving in CAD. It focused on automating routine tasks through predefined domain-specific knowledge. In contrast, ICAD 2.0 embraces a dynamic and interactive approach, leveraging advanced AI technologies to interpret diverse inputs and support a collaborative design process. This shift from rigid rules to adaptive learning enhances flexibility and responsiveness.

\textbf{Core Capabilities.}
ICAD 1.0 was designed to automate well-defined, repetitive tasks within narrow domains, excelling in efficiency and consistency. However, its reliance on symbolic reasoning limited its adaptability to complex and evolving design challenges. ICAD 2.0 extends these capabilities by incorporating multimodal AI, which processes interactive inputs such as sketches, voice commands, and gestures. The integration of generative AI and deep learning allows ICAD 2.0 to handle ambiguous inputs, predict outcomes, and suggest innovative alternatives, thus supporting both routine and complex design tasks more effectively.

\textbf{Key Techniques.}
ICAD 1.0 employed symbolic reasoning and domain-specific knowledge to automate tasks based on predefined rules. ICAD 2.0 introduces advanced techniques, including multimodal AI, which integrates graphical, verbal, and gestural inputs, and generative AI, which generates and refines design concepts. Deep learning provides predictive analytics and real-time insights, making the system more adaptive, intuitive, and comprehensive.

\textbf{Impact on the Design Process.}
ICAD 1.0 enhanced efficiency and consistency by automating repetitive tasks within well-defined domains. ICAD 2.0, on the other hand, enables a more interactive and adaptive design process. With real-time feedback and innovative recommendations, ICAD 2.0 facilitates a streamlined design process that integrates technical accuracy with creative exploration, resulting in faster iterations and improved design quality, especially in the early design stage.

\section{Conclusions}
\label{sec:conclusions}
In this paper, we have presented a vision of next-generation ICAD, defining its scope and requirements. By addressing the questions "Which activities in the design process can be managed by CAD and AI?" and "What are the needs of CAD, and how can AI contribute?", we argue that ICAD remains a human-computer collaborative system rather than an autonomous agent. Furthermore, ICAD should assume a more intensional role in the design process, meaning that it engages more deeply with what designers really think and actively manages tasks related to this process, particularly for the early design stages.

Based on the definition and requirements, we have also outlined a conceptual framework for the implementation of ICAD, including key components and potential technical challenges. By integrating recent advancements in machine learning, multimodal AI, and generative AI, we demonstrate how these technologies can be used to create a closed-loop among effective exchange of concepts between designers and computers, automated predictive analytics from both vague and precise design descriptions, and creative design refinement recommendations. With them, CAD can extend beyond merely detailed design stages into early design stages and even the integrated design process. 

\section*{Acknowledgement}
This work has been funded by NSF of China (No. 62102355), the ``Pioneer" and ``Leading Goose" R\&D Program of Zhejiang Province (No. 2024C01103), NSF of Zhejiang Province (No. LQ22F020012), and the Fundamental Research Funds for the Zhejiang Provincial Universities (No. 2023QZJH32).

The views presented in this paper are personal, but some stimulating discussions with Prof. Charlie C.L. Wang and Dr. Yang Liu have been invaluable.

 \bibliographystyle{elsarticle-num} 
 \bibliography{cas-refs}

\end{document}